\documentstyle[aps]{revtex}
\begin{document}

\title{Application of realistic effective interactions to the
structure of the Zr isotopes}
\author{A.\ Holt, T.\ Engeland, M.\ Hjorth-Jensen and E.\ Osnes}
\address{Department of Physics,
         University of Oslo, N-0316 Oslo, Norway}

\maketitle
\begin{abstract}
We calculate the low-lying spectra of the zirconium isotopes ($Z=40$) with
neutron numbers from $N=52$ to $N=60$ using the $1p_{1/2}0g_{9/2}$ proton and 
$2s1d0g_{7/2}0h_{11/2}$ neutron sub-shells to define the model space. Effective 
proton-proton, neutron-neutron and proton-neutron interactions have been 
derived using $^{88}$Sr as closed core and employing perturbative many-body 
techniques. The starting point is the nucleon-nucleon potential derived from 
modern meson exchange models. The comprehensive shell-model calculation
performed in this work provides a qualitative reproduction of essential 
properties such as the sub-shell closures in $^{96}$Zr and $^{98}$Zr.

\noindent {\it PACS number(s):\ } 21.60.Cs, 27.60.+j
\end{abstract}

\section{INTRODUCTION}

The Zr isotopes undergo a clear and smooth shape transition with increasing 
neutron number. The isotopes which are displayed in Figure \ref{fig:exp}(a) span 
from pure spherical nuclei that can be described in terms of simple 
shell-model configurations to the strongly deformed nucleus $^{102}$Zr. 
Evidence for coexisting shapes has been reported around $^{100}$Zr 
\cite{schu80,beck84,mach89,lher97}. In the intermediate region, both 
$^{96}$Zr and $^{98}$Zr 
present evidence for sub-shell closure. These isotopes have a remarkably large 
gap between the ground state and the $2^{+}_{1}$ level, $1.751$ MeV in 
$^{96}$Zr and $1.223$ MeV in $^{98}$Zr. They also have a 
relatively low level density below 3 MeV. Empirically, the Zr isotopes are
fairly well established. Lhersonneau and collaborators have 
recently performed careful experimental studies of the Zr isotopes and 
neighbouring nuclei, and they have made major contributions to the 
identification of levels in $^{97}$Zr and $^{99}$Zr \cite{lher94}.

For comparison the empirical Sr spectra are sketched in Figure \ref{fig:exp}(b).
The Sr isotopes differ from the Zr isotopes by only two protons but are 
qualitatively quite different. Compared to Zr the Sr isotopes have a much 
smoother behaviour with a stable $0^{+}-2^{+}$ spacing similar to that 
observed in tin. Large gaps due to sub-shell closure as observed in
$^{96}$Zr and $^{98}$Zr do not occur in Sr. However, as in Zr there is
a clear transition from spherical to deformed shape around $N=60$. The 
low-lying spectrum of $^{100}$Sr is a nearly perfect rotational band.
It is a theoretical challenge to describe a sequence of isotopes with 
such big changes in the structure from one nucleus to another as in Zr. 
For a proper description of the Zr isotopes one has to allow for proton 
excitations. In particular, the protons seem to play a dominant role in the 
$0^{+}_{2}$ state. Thus, the common choice of inert core has been $^{88}$Sr, 
though with large variations in the size of model space, truncation scheme 
and effective interactions, Refs. \cite{gloe75,hals93,ips75}.
The early calculation by Auerbach and Talmi \cite{auer65} was carried out 
with valence protons filling the ($1p_{1/2}, 0g_{9/2}$) oribitals and
valence neutrons filling the ($1d_{5/2}$) orbital. In particular if one wants
to describe more neutron rich Zr isotopes a larger model space is required
as the interaction between ($0g_{9/2}$) protons and ($2d_{3/2}$) and ($0g_{7/2}$)
neutrons becomes increasingly important \cite{gloe75}. To our knowledge the 
present work is the first shell-model calculation of the Zr-isotopes which 
includes nuclei up to $^{98}$Zr within a non-truncated $p:(1p_{1/2},0g_{9/2})$ 
$n:(1d_{5/2},2s_{1/2},1d_{3/2},0g_{7/2},0h_{11/2})$ model space and with a 
fully realistic effective interaction.

We will in the present work perform a systematic shell-model study of the 
Zr isotopes from N=50 to N=58, and will in particular pay attention to
the nuclei around the closure of the neutron $(1d_{5/2})$ and
$(2s_{1/2})$ sub-shells, $^{96-98}$Zr. 
In several works we have performed thorough analyses of the effective 
two-body interaction. We have derived shell-model effective 
interactions based on meson exchange models for the free nucleon-nucleon 
interaction, using many-body perturbation theory as described below. The 
systems that have been studied are reaching from the oxygen 
region to the tin isotopes and the N=82 isotones, Refs. 
\cite{eng93,holt97,suh98,holt98}. For the lighter systems, such as the $sd$-
and $pf$-shell nuclei, we obtained markedly better results for nuclei with 
one kind of valence nucleons than for nuclei with both kinds. For the heavier
systems we have so far restricted ourselves to nuclei with like valence 
nucleons, such as the Sn isotopes and the $N=82$ isotones. This way we have
managed to keep the dimensionality of the eigenvalue problem within tractable
limits. Further, we have seen the need for establishing confidence in the
$T=1$ interaction before considering systems with both valence protons and
neutrons, where the proton-neutron interaction may play a crucial role. In
fact, the Zr isotopes represent a challenge on both these accounts.

Let us in terms of a simplified model like the weak coupling scheme, in which 
the proton-neutron interaction is assumed to be weak, see if one 
can gain insight into the qualitative properties of this interaction. In
Figure \ref{fig:wc} we demonstrate the validity of this scheme by seeing how well
it describes properties of $^{92,94,96}$Zr. In column one the 
empirical spectrum of $^{90}$Zr (dashed lines), which represents the proton 
degrees of freedom, is put on top of the $^{90}$Sr spectrum (solid lines), which 
represents the neutron degrees of freedom. This would represent the $^{92}$Zr 
spectrum in the weak coupling limit and should be compared with the empirical 
$^{92}$Zr spectrum in column two. Similarly the weak coupling spectra $^{92}
{\rm Sr} + ^{90}$Zr and $^{94}{\rm Sr} + ^{90}$Zr are compared with
the empirical $^{94}$Zr and  $^{96}$Zr 
spectra, respectively. Most states in $^{92}$Zr and $^{94}$Zr are well 
reproduced by the weak coupling scheme. The model does however collapse in 
$^{96}$Zr, due to the presumed closure of the $1d_{5/2}$ sub-shell which does 
not have a counterpart in $^{94}$Sr. The fact that the weak coupling scheme is 
fairly successful in describing $^{92}$Zr and $^{94}$Zr leads us to believe 
that the proton-neutron part of the effective interaction is either rather
weak or state-independent.

The paper is organized as follows. In Sect. 2 we give a summary of the 
calculation of the effective interaction and the shell model. Then
the results are presented and discussed in Sect. 3 and conclusions are drawn 
in Sect. 4.

\section{THEORETICAL FRAMEWORK}

The aim of microscopic nuclear structure calculations is to derive various 
properties of finite nuclei from the underlying hamiltonian describing the 
interaction between nucleons. When dealing with nuclei, such as the Zr 
isotopes with $A = 90 - 100$, the full dimensionality of the many-body 
Schr\"{o}dinger equation
\begin{equation}
     H\Psi_i(1,...,A)=E_i\Psi_1(1,...,A),
     \label{eq:full_a}
\end{equation}
becomes intractable and one has to seek viable approximations to 
Eq.\ (\ref{eq:full_a}). In Eq.\ (\ref{eq:full_a}), $E_i$ and $\Psi_i$ 
are the eigenvalues and eigenfunctions for a state $i$ in the Hilbert space.

One is normally only interested in solving Eq.\ (\ref{eq:full_a})
for certain low-lying states. It is then customary to divide the Hilbert 
space into a model space defined by the operator $P$ and an excluded space
defined by a projection operator $Q=1-P$
\begin{equation}
     P=\sum_{i=1}^{d}\left | \psi_i\right\rangle 
     \left\langle\psi_i\right | \hspace{1cm}    
 Q=\sum_{i=d+1}^{\infty}\left | \psi_i\right\rangle 
     \left\langle\psi_i\right | ,
\end{equation}
with $d$ being the size of the model space and such that $PQ=0$.
The assumption is that the low-lying states can be fairly well reproduced by 
configurations consisting of a few particles occupying physically selected 
orbitals, defining the model space. In the present work, the model space to 
be used both in the shell-model calculation and in the derivation of the effective 
interaction is given by the proton orbitals $1p_{1/2}$ and $0g_{9/2}$ and the 
neutron orbitals $2s_{1/2}$, $1d_{5/2}$, $1d_{3/2}$, $0g_{7/2}$ and $0h_{11/2}$.

Eq.\ (\ref{eq:full_a})  can then be rewritten as a secular equation
\begin{equation}
    PH_{\mathrm{eff}}P\Psi_i=P(H_{0}+V_{\mathrm{eff}})
    P\Psi_i=E_iP\Psi_i,
\end{equation}
where $H_{\mathrm{eff}}$  now is an effective hamiltonian acting solely
within the chosen model space. The term $H_0$
is the unperturbed hamiltonian while the effective interaction is
given by
\begin{equation}
  V_{\mathrm{eff}}=\sum_{i=1}^{\infty} V_{\mathrm{eff}}^{(i)},
\end{equation}
with $ V_{\mathrm{eff}}^{(1)}$,  $ V_{\mathrm{eff}}^{(2)}$,
 $ V_{\mathrm{eff}}^{(3)}$,...\ being effective one-body, two-body,
three-body interactions etc. 
It is also customary in nuclear shell-model calculations to add
the one-body effective interaction  $ V_{\mathrm{eff}}^{(1)}$
to the unperturbed part of the hamiltonian so that
\begin{equation}
    H_{\mathrm{eff}}=\tilde{H}_{0}+  V_{\mathrm{eff}}^{(2)}+
    V_{\mathrm{eff}}^{(3)}+\dots,
\end{equation}     
where $\tilde{H}_{0}=H_{0}+V_{\mathrm{eff}}^{(1)}$. This allows us
 to replace the eigenvalues of 
$\tilde{H}_{0}$ by the empirical single-particle energies 
for the nucleon orbitals of our model space, or valence space.
Thus, the remaining quantity to calculate is the two- or more-body
effective interaction 
$\sum_{i=2}^{\infty} V_{\mathrm{eff}}^{(i)}$.
In this work we will restrict our attention to the derivation of
an effective two-body interaction 
\begin{equation}
      V_{\mathrm{eff}}=V_{\mathrm{eff}}^{(2)},
\end{equation}
using the many-body methods discussed in Ref.\ \cite{hko95}
and briefly reviewed below.

\subsection{Effective interaction}

Our procedure for obtaining  an effective interaction for 
the Zr isotopes
starts with a free nucleon-nucleon  interaction $V^{(2)}$ which is
appropriate for nuclear physics at low and intermediate energies. 
At present, there are several potentials available. The most recent 
versions of Machleidt and co-workers
\cite{cdbonn}, the Nimjegen group \cite{nim} and the Argonne
group \cite{v18} have all a $\chi^2$ per datum close to $1$.
In this work we will thus choose to work with the charge-dependent
version of the Bonn potential models, see Ref.\ \cite{cdbonn}.
The potential model of Ref.\ \cite{cdbonn} is an extension of the 
one-boson-exchange models of the Bonn group \cite{mac89}, where mesons 
like $\pi$, $\rho$, $\eta$, $\delta$, $\omega$ and the fictitious
$\sigma$ meson are included. In the charge-dependent version
of Ref.\ \cite{cdbonn}, the first five mesons have the same set
of parameters for all partial waves, whereas the parameters of
the $\sigma$ meson are allowed to vary. 

The next step in our perturbative many-body scheme is to handle 
the fact that the strong repulsive core of the nucleon-nucleon potential $V$
is unsuitable for perturbative approaches. This problem is overcome
by introducing the reaction matrix $G$ given by the solution of the
Bethe-Goldstone equation
\begin{equation}
    G=V+V\frac{Q}{\omega - QTQ}G,
\end{equation}
where $\omega$ is the unperturbed energy of the interacting nucleons,
and $H_0$ is the unperturbed hamiltonian. 
The operator $Q$, commonly referred to
as the Pauli operator, is a projection operator which prevents the
interacting nucleons from scattering into states occupied by other nucleons.
In this work we solve the Bethe-Goldstone equation for five starting
energies $\omega$, by way of the so-called double-partitioning scheme
discussed in e.g.,  Ref.\ \cite{hko95}. 
The $G$-matrix is the sum over all
ladder type of diagrams. This sum is meant to renormalize
the repulsive short-range part of the interaction. The physical interpretation
is that the particles must interact with each other an infinite number
of times in order to produce a finite interaction. 

Finally, we briefly sketch how to calculate an effective 
two-body interaction for the chosen model space
in terms of the $G$-matrix.  Since the $G$-matrix represents just
the summmation to all orders of ladder diagrams with particle-particle
intermediate states, there are obviously other terms which need to be included
in an effective interaction. Long-range effects represented by 
core-polarization terms are also needed.
The first step then is to define the so-called $\hat{Q}$-box given by
\begin{equation}
   P\hat{Q}P=PGP +
   P\left(G\frac{Q}{\omega-H_{0}}G\\ + G
   \frac{Q}{\omega-H_{0}}G \frac{Q}{\omega-H_{0}}G +\dots\right)P.
   \label{eq:qbox}
\end{equation}
The $\hat{Q}$-box is made up of non-folded diagrams which are irreducible
and valence linked. A diagram is said to be irreducible if between each pair
of vertices there is at least one hole state or a particle state outside
the model space. In a valence-linked diagram the interactions are linked
(via fermion lines) to at least one valence line. Note that a valence-linked
diagram can be either connected (consisting of a single piece) or
disconnected. In the final expansion including folded diagrams as well, the
disconnected diagrams are found to cancel out \cite{ko90}.
This corresponds to the cancellation of unlinked diagrams
of the Goldstone expansion \cite{ko90}.
These definitions are discussed in Refs.\ \cite{hko95,ko90}.
We can then obtain an effective interaction
$H_{\mathrm{eff}}=\tilde{H}_0+V_{\mathrm{eff}}^{(2)}$ in terms of the 
$\hat{Q}$-box \cite{hko95,ko90}, with 
\begin{equation}
    V_{\mathrm{eff}}^{(2)}(n)=\hat{Q}+{\displaystyle\sum_{m=1}^{\infty}}
    \frac{1}{m!}\frac{d^m\hat{Q}}{d\omega^m}\left\{
    V_{\mathrm{eff}}^{(2)}(n-1)\right\}^m,
    \label{eq:fd}
\end{equation}
where $(n)$ and $(n-1)$ refer to the effective interaction after
$n$ and $n-1$ iterations. The zeroth iteration is represented by just the 
$\hat{Q}$-box.
Observe also that the
effective interaction $V_{\mathrm{eff}}^{(2)}(n)$
is evaluated at a given model space energy
$\omega$, as is the case for the $G$-matrix as well. Here we choose
$\omega =-20$ MeV.
Moreover, although $\hat{Q}$ and its derivatives contain disconnected
diagrams, such diagrams cancel exactly in each order \cite{ko90}, thus
yielding a fully connected expansion in Eq.\ (\ref{eq:fd}).
Less than $10$ iterations were needed in order to obtain a numerically
stable result. All non-folded diagrams through 
third order in the interaction $G$ are included.
For further details, see Ref.\ \cite{hko95}.

\subsection{Shell model}

The effective two-particle interaction can in turn be used in shell-model
calculations.
Our approach in solving the eigenvalue problem
is the Lanczos algorithm, which is an iterative method that gives the solutions
of the lowest eigenstates. The technique is described in detail in Refs.
\cite{whi77,eng95}.
The shell-model code developed by us is designed for an $m$-scheme Slater 
determinant (SD) representation. Even with a rather restricted 
single-particle basis the size of the shell-model problem grows rapidly with 
increasing number of active valence particles. 
Table \ref{tab:dimension} shows how the number of configurations in an 
SD-basis grows with the number of valence particles acting within the 
($1p_{1/2}0g_{9/2}$) proton shell and the ($2s1d0g_{7/2}0h_{11/2}$) neutron 
shell relative to the $^{88}$Sr-core. Note that the $^{98}$Zr system consists
of more than 26.000.000 basis states. Only a few years ago shell-model 
calculations on such systems were not tractable. Even with today's
fast computers and effective algorithms this kind of calculation is
still rather time consuming.

The single-neutron energies are taken to be those deduced from $^{89}$Sr
in Refs. \cite{saga84,clea78}. In the literature the single-proton energy 
splitting $\varepsilon (1p_{1/2}) -  \varepsilon (0g_{9/2})$ varies from 
0.839 MeV to 1.0 MeV \cite{hals93,ips75,rama73,hoff74,chuu79}. As the final 
results show little sensitivity to variations within this energy interval we 
let the $0g_{9/2}$ single-proton energy relative to $1p_{1/2}$ be 0.9 MeV. 
The single-particle energies used in this work are listed in Table \ref{tab:sp}.

\section{RESULTS AND DISCUSSIONS}

In this section we present the calculations of the Zr isotopes. 
Firstly, we analyze some systematics of the even Zr isotopes. The
results are displayed in Figures \ref{fig:92-94zr} and \ref{fig:96-98zr} 
(for some selected states) and in  more detail
in Tables \ref{tab:90-92zr}, \ref{tab:94-96zr} and \ref{tab:98zr}.
Then, we proceed with the discussion of the odd Zr isotopes, in Figure 
\ref{fig:97zr} and Tables \ref{tab:91-93zr} and \ref{tab:95-97zr}. A major
aim is to investigate the effective interaction that has been derived for 
this mass region. Since the odd nuclei are generally more sensitive to the 
underlying assumptions made, this may give an even more severe test of the 
interaction and the foundations on which our model is based. At the end of
this section we discuss problems concerning the binding energies.
In order to study the effect of the proton degrees of freedom we have also
performed a more restricted calculation, considering only valence neutrons 
with respect to a $^{90}$Zr-core.

\subsection{Even isotopes}

The experimental spectra of the two nuclei, $^{92}$Zr and $^{94}$Zr, show
very similar features. The shell-model calculation does also provide 
$0^{+}_{1}$, $2^{+}_{1}$ and $4^{+}_{1}$ levels in $^{92}$Zr and $^{94}$Zr
with quite similar features, although the three levels are too compressed 
compared to their experimental counterparts. Comparison of the results 
obtained with $^{90}$Zr- and $^{88}$Sr-cores indicates that these levels 
are little affected by proton excitations. In $^{96}$Zr the calculated 
$0^{+}_{1}$, $2^{+}_{1}$ and $4^{+}_{1}$ levels are still too compressed, 
though not as pronounced as in $^{92}$Zr and $^{94}$Zr, while in $^{98}$Zr 
the calculated spectrum is more open than the experimental one.

The energy of the empirical $3^{-}_{1}$ level, Figure \ref{fig:exp}(a), is 
monotonously reduced with increasing neutron number. In $^{92}$Zr and  $^{94}$Zr 
the calculated $3^{-}_{1}$ level is obtained at about 2 MeV, while in $^{96}$Zr 
and  $^{98}$Zr the $3^{-}_{1}$ level is obtained much too high at about 3.5 MeV. 
Due to the extension of the model space from a $^{90}$Zr-core to a 
$^{88}$Sr-core the $3^{-}_{1}$ level in  $^{92}$Zr and $^{94}$Zr undergoes a 
considerable lowering, yielding results close to the experimental values. From 
the occupation numbers in Table \ref{tab:occupation} it is clear that the 
$3^{-}_{1}$ state undergoes a structural change  from $^{94}$Zr to
$^{96}$Zr. The $0g_{7/2}$ and $2s_{1/2}$ orbitals
start to play a more important role. For comparison, the observed 
$3^{-}_{1}$ levels in Sr, Figure \ref{fig:exp}(b), are all situated around
2 MeV. Their calculated 
counterparts are located too high, at about 3 MeV excitation energy. The 
structure of the $3^{-}_{1}$ level in Zr is totally different from the structure
in Sr, Figure \ref{fig:sr}. Let us look in detail into these states by comparing 
the $3^{-}_{1}$ levels in $^{92}$Sr and $^{94}$Zr. Both nuclei have four 
valence neutrons. In $^{92}$Sr the dominant configuration is 
$[(1d_{5/2})^{3}0h_{11/2}]_{J=3^{-}}$ whereas in $^{94}$Zr the dominant 
configuration is 
$[(1p_{1/2}0g_{9/2})_{J_{\pi}=4^{-},5^{-}}(1d_{5/2})^{4}_{J_{\nu}=2^{+}}]_{J=3^{-}}$.
The difference can be ascribed to the single-particle energies. 
In Zr the $3^{-}_{1}$ state is created by exciting a proton into the $0g_{9/2}$
orbital instead of exciting a neutron into the $0h_{11/2}$ orbital. Because the
$0h_{11/2}$ orbital is located very high in the single-particle spectrum it is 
more favourable to excite a proton rather than a neutron.  

The situation for the $5^{-}_{1}$ level is more stable throughout the whole
sequence of Zr isotopes, with empirical values between 2.5 and 3.0 MeV. The
$^{90}$Zr-core calculations provide level energies that are 1.0 - 1.5 MeV too
high, while the $^{88}$Sr-core calculations give energies too low by about 
1 MeV. This state consists predominantly of configurations with a proton 
excited into the $0g_{9/2}$ orbital and the neutrons remaining in the lowest 
possible single-particle orbitals.

As pointed out in the introduction, there are strong variations in the 
structure from one nucleus to another, reflected in for instance the 
$0^{+}-2^{+}$ spacing. Qualitatively we reproduce the variation in the 
$0^{+}-2^{+}$ spacing quite well, as shown in Figure \ref{fig:2-pluss}, 
although in $^{90-96}$Zr the gap is $200 - 400$ keV less than the experimental 
spacing. 

In spite of clear differences in the experimental Zr and Sr spectra, the 
shell-model calculation provides rather similar results. The calculated Zr 
spectra are in far better agreement with experimental data than the 
calculated Sr spectra, which may indicate that the core is in a
different condition in the two systems. In Sr the core seems to be
relatively soft, whereas in Zr the two additional protons tend to
stabilize the core.

\subsubsection{Proton configurations}
\label{sec-protonconfig}

The proton degrees of freedom are crucial in order to describe certain energy 
levels. For example, the first excited $0^{+}_{2}$ state in $^{92}$Zr has strong 
components of proton excitations. From the occupation numbers in 
Table \ref{tab:occupation0} we see that the character of $0^{+}_{1}$ state is
totally different from the $0^{+}_{2}$ state. The proton parts of their 
wave functions are almost orthogonal to each others. In the 
ground state the protons are most likely to be found in the 
$1p_{1/2}$-orbital, while for the excited $0^{+}$-state it is more probable 
to find the protons in the $0g_{9/2}$-orbital. However, in $^{96}$Zr the two 
shell-model $0^{+}$ states have nearly the same proton structure, almost pure 
$(1p_{1/2}^{2})_{\pi}$ configuration. The $0^{+}$ levels in $^{98}$Zr show similar 
structure as in $^{96}$Zr, but the $0^{+}_{2}$ state has slightly stronger 
$0g_{9/2}$ mixing than in $^{96}$Zr. The third $0^{+}$ state in both $^{96}$Zr and
$^{98}$Zr have $(0g_{9/2})$ as the predominant proton configuration. The 
change in the proton configuration with increasing neutron 
number was observed in pick-up experiments by Saha {\em et al.} 
\cite{saha79}. 
\subsubsection{Sub-shell closure}

Clear signs of sub-shell closure are seen in $^{96}$Zr and also in $^{98}$Zr,
due to filling of the $1d_{5/2}$ and the $2s_{1/2}$ orbitals, respectively.
From $^{94}$Zr to $^{96}$Zr the gap between the ground state and the 
$2^{+}_{1}$ state is doubled. The experimental $0^{+}_{1} \rightarrow 
2^{+}_{1}$ gap increases, from 0.919 MeV in $^{94}$Zr to 1.751 MeV in 
$^{96}$Zr, and the calculated gap is also more than doubled, from 0.520 MeV 
to 1.426 MeV. In $^{98}$Zr the calculated spacing is larger than the 
experimental one. The experimental $0^{+}-2^{+}$ spacing is 1.223 MeV, and 
the corresponding calculated spacing 1.463 MeV. The $^{96}$Zr ground state 
$1d_{5/2}$ occupation number is 5.66, and in $^{98}$Zr the $1d_{5/2}$ and
$2s_{1/2}$ occupation numbers are 5.76 and 1.87, respectively.

The total impression of the $^{98}$Zr shell-model results displayed in Figure 
\ref{fig:96-98zr} is disappointing. Only the $2^{+}_{1}$ level is reasonably 
reproduced. As an alternative to the extremely time and space consuming  
calculation presented in Figure \ref{fig:96-98zr}, we may close the $1d_{5/2}$ 
orbital and perform a $^{94}$Sr-core shell-model calculation of the system. 
The results, shown in Table \ref{tab:98zr}, are much improved.

\subsubsection{E2 transition rates}

Experimental and calculated E2 transition rates are tabulated in 
Table \ref{tab:e2}.
In order to bring the theoretical results into agreement with the measured 
$2^{+}_{1}\longrightarrow 0^{+}_{1}$ transition rates we have employed 
effective charges of 1.8$e$ and 1.5$e$ for the protons and neutrons, 
respectively.
These values are consistent with the effective charges obtained in Refs. 
\cite{gloe75,chuu79,brow76}, however a bit overestimated compared to the 
fitting to data on $^{92}$Mo and $^{91}$Zr done by Halse in Ref. \cite{hals93}.  
To be mentioned, the calculation of effective charges based on perturbative 
many-body methods \cite{hjor99} gives much smaller values, $1.1e$ and
in the range $0.5e - 0.7e$ for the proton and neutron effective
charge, respectively.

We adopt Halse's effective value of the oscillator parameter $b=2.25$ fm. The 
value was choosen by reference to measurements for the radii of the 
single-particle orbitals in $^{89}$Sr and of the charge 
distributions in $^{92-96}$Mo, Refs. \cite{saga84} and \cite{sche80}.

With effective charges and the oscillator parameter as described above, the 
transition rates between yrast states are fairly well reproduced. In the 
former discussion we have focused on the $0^{+}_{2}$ state, in particlular 
its proton structure. In $^{92}$Zr and $^{94}$Zr we totally fail in 
reproducing the transition rates involving the $0^{+}_{2}$ state. The 
experimental transition rates between $0^{+}_{2}$ and $2^{+}_{1}$ in 
$^{92}$Zr and $^{94}$Zr are relatively strong, 14.3(5) and 9.3(4) W.u., 
respectively, whereas the calculated transition rates are two to three
orders of magnitude smaller.
Similarly, in $^{98}$Zr there is an experimental transition rate between 
$0^{+}_{3}$ and $2^{+}_{1}$ with strength 51(5) W.u. The corresponding 
calculated transition rate is negligible. In fact, the occupation 
numbers in Table \ref{tab:occupation0} show that the $0^{+}_{3}$ state in 
$^{98}$Zr has similar proton structure to the $0^{+}_{2}$ state in $^{92}$Zr 
and $^{94}$Zr. Leaving out the neutron contributions, as done in column 6 of 
Table \ref{tab:e2}, by setting the effective neutron charge equal to zero, 
we observe that the proton part of the wave function contributes more to the
transitions involving the excited $0^{+}$ states than what it does to the 
other transition rates. The contribution is however far from sufficient and 
there is a cancellation effect between the proton and neutron contributions.
Contributions to the transition rates between yrast states do on the other
hand mainly stem from the neutron degrees of freedom. 
In conclusion, it seems that the $0^{+}_{2}$ states in $^{92}$Zr and
$^{94}$Zr and the $0^{+}_{3}$ state in $^{98}$Zr contain strong
collective components not reproduced by the present shell model calculation.

\subsection{Odd isotopes}

It is somehow surprising to notice that the shell model gives a much better 
description of the odd than of the even Zr isotopes. The reproduction of the 
low-lying positive parity states are overall satisfactory. On the other 
hand the shell model has some problems in describing the negative parity 
states. Several of the negative parity states in $^{91,93}$Zr are calculated 
up to 1 MeV too low. Only a few negative parity states are known in $^{95,97}$Zr.
Thus a detailed comparison is difficult, but the $11/2^{-}_{1}$ state is 
reproduced in nice agreement with experiment. 

We will make a detailed study of $^{93}$Zr. This nucleus is not too 
simple and not too complex (two protons and three neutrons outside the 
closed core), and useful information can therefore be extracted from a 
few central and relatively simple configurations.

The shell-model calculation provides three states below 600 keV, 
$5/2_{1}^{+}$, $3/2_{1}^{+}$ and $9/2_{1}^{+}$. From experiments, only two 
states are known, $5/2_{1}^{+}$ and $3/2_{1}^{+}$. There are
however theoretical arguments supporting a low-lying $9/2_{1}^{+}$-state.
The configuration that requires the least energy has all particles 
in the lowest possible single-particle orbital. In the case of $^{93}$Zr the 
two protons occupy the $1p_{1/2}$ single-particle orbital, and couple to 
$J=0$. The  three neutrons occupy the $1d_{5/2}$-orbital 
$(1d_{5/2}^{3})_{\nu}$. The three neutrons can then couple to 
$J^{\pi}=3/2^{+}$, $5/2^{+}$ or $9/2^{+}$. Such states will be located well 
below 1 MeV. The lowest experimental $9/2^{+}$ candidate observed up to now 
is seen at 1.46 MeV.

As many as four $1/2^{+}$ states are observed within a small energy interval
of 250 keV in the region from 0.95 MeV to 1.22 MeV. This observation has no 
shell-model counterpart. Our calculation provides only one $1/2^{+}$ level
at 1.40 MeV.

We already pointed out that our model has difficulty in describing the 
negative parity states. Consider for example the structure of the 
$11/2^{-}_{1}$ state in $^{93}$Zr, which comes 500 keV
lower than the experimental position. Odd parity states are constructed
by configurations with an odd number of particles in negative parity states.
Within our model, protons can occupy the $1p_{1/2}$-orbital and neutrons can
occupy the $0h_{11/2}$-orbital to produce negative parity states.
Both the first and second excited $11/2^{-}$ states in $^{93}$Zr are 
predominantly based on the proton configuration $(1p_{1/2}0g_{9/2})_{\pi}$.
The same is true for the $11/2^{-}_{1}$ state in the other Zr isotopes.

Finally, we examine $^{97}$Zr. As we would expect for a system caught in 
between two ``magic'' nuclei we recognize pronounced  single-particle structure.
From the occupation numbers, listed in Table \ref{tab:occ}, we see
that both the ground state, $1/2^{+}_{1}$, and the next level, $3/2^{+}_{1}$, 
are pure one-quasiparticle states built on a full $1d_{5/2}$-orbital, i.e. 
$^{96}$Zr-core.
Also the $7/2^{+}_{1}$ state is a one-quasiparticle state with the 
$1d_{5/2}$-orbital nearly closed, though its calculated position is about 0.7 
MeV too high. This means that a somewhat lower single-particle energy 
$\varepsilon _{0g_{9/2}}$might be more appropriate for our effective interaction.
The $5/2^{+}_{1}$ state can be regarded as a $1d_{5/2}$-hole 
state relative to the $^{98}$Zr-core. 
All in all the yrast states apart from $7/2^{+}$ are very
well reproduced. In the other, non-yrast states, the $^{96}$Zr-core breaks up
and two neutrons are distributed equally among the $2s_{1/2}$ and $1d_{3/2}$ 
orbitals.

\subsection{Binding energies}

The binding energies calculated by the formula
\begin{eqnarray}
  {\rm BE}(^{90+n}{\rm Zr}) & = & {\rm BE}(^{90+n}Zr) - {\rm BE}(^{88}{\rm Sr}) \nonumber \\
                            & - & n({\rm BE}(^{89}{\rm Sr})-{\rm BE}(^{88}{\rm Sr})) \nonumber \\
                            & - & 2({\rm BE}(^{89}{\rm Y})-{\rm BE}(^{88}{\rm Sr})) \label{eq:be}
\end{eqnarray}
are plotted in Figure \ref{fig:be}. In Eq.\ (\ref{eq:be}) $n$ is the number of
valence neutrons.
The experimental binding energies show a parabola 
structure with a minimum at $^{96}$Zr, whereas the calculated binding energies 
increase linearly down to $^{98}$Zr. With increasing  neutron number the systems 
become far too strongly bound. This phenomenon of overbinding 
of nuclear systems when effective interactions from meson theory are
used has been much discussed in the literature, for example in Ref. 
\cite{zuk94}. The solutions to the problem has been that such matrix
elements must be modified in order to reproduce the binding energies 
correctly. The so-called centroid matrix elements should be corrected 
in order to reproduce experiment. However, there is no well-defined 
recipe for doing this. 

The curve labeled ``no pn-int'' in Figure \ref{fig:be} shows
the binding energies with the proton-neutron interaction switched off. Now,
the systems are too weakly bound, which tells us that the proton-neutron part 
of the interaction contributes strongly to the overbinding.
We have made the pn-interaction less attractive by adding an overall constant
to the diagonal matrix elements. This will not affect the excitation energies 
relative to the ground state. The constant is chosen such as to fit the 
experimental binding energy of $^{90}$Y. Thus, a constant 0.3 MeV added to the 
original diagonal proton-neutron matrix elements, 
$V_{\rm abab}^{\rm mod}({\rm pn}) = V_{\rm abab}^{\rm eff}({\rm pn}) + 0.3 
{\rm MeV}$, 
gives binding energies as shown in  Figure \ref{fig:be}, labeled ``modified
pn-int.''. The fit to the experimental values are much improved, although there 
is a linear rather than parabolic dependence on the particle number.

\section{CONCLUSIONS}

In this work we have performed a full $(1p_{1/2}0g_{9/2})$ proton and 
$(2s1d0g_{7/2}0h_{11/2})$ neutron shell-model calculation of the zirconium
isotopes ranging from N=52 to N=60. For the first time we present results
from calculations with a proton-neutron effective interaction in such heavy
nuclei.

We have succeeded in obtaining a qualitative reproduction of important 
properties, 
although there are also shortcomings. The odd isotopes are very well described 
by our shell model, in fact better than the calculated even isotopes. Both for 
the odd and the even nuclei we have difficulties in reproducing  the negative 
parity states well. 

For comparison we have presented shell-model results for the neighbouring 
strontium isotopes. The qualitative features are much better reproduced in Zr 
than in Sr. For example, the shell model fails in reproducing the very stable 
$0^{+} - 2^{+}$ spacing in Sr. Differently from Zr, there is no sign of
N=56 and N=58 sub-shell closures in Sr. The quality of the closed-shell core 
($^{88}$Sr) may in fact be different for the two sets of isotopes.
It is likely that the additional protons in Zr give a more balanced system and 
serve to stabilize the core. In Sr the core seems to be softer and more unstable.

The empirical Zr spectra can to a certain extent be interpreted in a weak 
coupling scheme. The isotopes $^{92}$Zr and  $^{94}$Zr are well described 
in terms of this model, which in turn indicates that the 
proton-neutron interaction should not be too strong. The simple weak coupling
picture collapses however in $^{96}$Zr.

In order to obtain results for $^{98}$Zr, we performed calculations that are 
extremely heavy and time consuming. All the efforts gave final results that 
were far off, and in fact a much simpler calculation based on a 
$^{94}$Sr-core provided results in better agreement with the experimental data.

The occupation numbers give a hint that the $0g_{7/2}$ and the $0h_{11/2}$
neutron orbitals are of minor importance. The major properties are in fact fairly
well described within a reduced basis $(1p_{1/2},0g_{9/2})_{\pi}$ 
$(2s_{1/2},1d_{5/2},1d_{3/2})_{\nu}$. 

In order to further test the wave functions we calculated E2
transition rates in the even Zr isotopes. Transitions between yrast
states are fairly well reproduced, whereas transitions involving
certain excited $0^{+}$ states are calculated far too small,
indicating that these states contain strong components not 
accounted for by the present shell model.

As in other mass regions we fail in reproducing bulk properties such as the
binding energies. With increasing number of valence nucleons the systems become
far too strongly bound. We have demonstrated that this problem can be cured by 
simple adjustments of some selected matrix elements.

The calculations have been carried out at the IBM cluster at 
the University of Oslo. Support for this from the Research
Council of Norway (NFR) is acknowledged. 
\appendix
\section{Tables}
The low-lying experimental and calculated energy-levels in
$^{90-98}$Zr are listed in Tables
\ref{tab:90-92zr}, \ref{tab:94-96zr} and \ref{tab:98zr}.
%

%
 \begin{figure}
\caption{\label{fig:exp} Experimental low energy level schemes for 
the Zr (a) and Sr (b) isotopes from N=50 to N=62.}
 \end{figure}
\bigskip
 \begin{figure}
\caption{\label{fig:wc}Demonstration of the weak coupling scheme. The
$^{90}$Zr energy levels are represented by dashed lines. Experimental
values are used for all levels energies.}
 \end{figure}
\bigskip
 \begin{figure}
\caption{\label{fig:92-94zr}Selected energy levels in $^{92}$Zr and  $^{94}$Z. 
Numbers in parentheses are the binding energies of the valence nucleons 
outside a $^{88}$Sr-core.}
 \end{figure}
\bigskip
 \begin{figure}
\caption{\label{fig:96-98zr} Selected energy levels in $^{96}$Zr and  
$^{98}$Z. Numbers in parentheses are the binding energies of the valence 
nucleons outside a $^{88}$Sr-core.}
\end{figure}
\bigskip
 \begin{figure}
 \caption{\label{fig:sr}Calculated energy levels for the Sr isotopes from
N=52 to N=60. The experimental levels are shown in Fig. 1b.}
 \end{figure}
\bigskip
 \begin{figure}
 \caption{\label{fig:2-pluss} The $2^{+}_{1}$ excitation energy, 
experimental (solid) and calculated, complete (dashed) and 
without pp- and pn-interaction (dotted), respectively.}
 \end{figure}
\bigskip
 \begin{figure}
\caption{\label{fig:97zr}Theoretical and experimental spectra of $^{97}$Zr.}
 \end{figure}
\bigskip
 \begin{figure}
 \caption{\label{fig:be}Relative binding energies of $^{90}$Zr -
 $^{98}$Zr. The curves show the experimental and theoretical binding
 energy curves, the full shell-model calculation, calculations 
 without the pn-interaction and calculations with modified 
 pn-interaction, respectively.}
 \end{figure}
%
\begin{table}[htbp]
\begin{center}
\caption{Number of shell-model basis states in $m$-scheme SD representation.}
\begin{tabular}{cr|cr|cr}
$n_{p} + n_{n}$ & Dimension & $n_{p} + n_{n}$ & Dimension  & $n_{p} + n_{n}$ & Dimension\\
           \hline
2 + 0 &       8  & 2 + 3 &  15 868 & 2 + 6  &  2 428 814 \\
2 + 1 &     186  & 2 + 4 & 107 060 & 2 + 7  &  8 648 777 \\ 
2 + 2 &   1 572  & 2 + 5 & 564 393 & 2 + 8  & 26 201 838 \\
\end{tabular}
\label{tab:dimension}
\end{center}
\end{table}

\begin{table}[htbp]
\begin{center}
\caption{Single-particle energies, all entries in MeV.}
\begin{tabular}{lll|llllll}
\multicolumn{3}{c|}{Single-proton energies} & \multicolumn{6}{c}{Single-neutron energies} \\
$j_{n}$    & $1p_{1/2}$ & $0g_{9/2}$ & 
\hspace{.4cm} $j_{n}$ & $1d_{5/2}$ & $2s_{1/2}$ 
& $1d_{3/2}$ & $0g_{7/2}$ & 
$0h_{11/2}$ \\
\hline
$\varepsilon (j_{n})$ & 0.00 & 0.90 &  \hspace{.4cm} $\varepsilon (j_{n})$ & 0.00 & 1.26 & 2.23 & 2.63 & 3.50 \\
\end{tabular}
\label{tab:sp}
\end{center}
\end{table}
\begin{table}[htbp]
\begin{center}
\caption{Occupation numbers of the $2^{+}_{1}$, $4^{+}_{1}$, $3^{-}_{1}$
and $5^{-}_{1}$ states in $^{92-98}$Zr.}
\begin{tabular}{ll|ll|lllll}
   & J$^{\pi}_{i}$ & $0g_{9/2}$ & $1p_{1/2}$ & $0h_{11/2}$ & $0g_{7/2}$ &
                $1d_{5/2}$ & $1d_{3/2}$ & $2s_{1/2}$ \\
\hline
$^{92}$Zr: &             &      &      &      &      &      &      &      \\
           & $2^{+}_{1}$ & 0.27 & 1.73 & 0.02 & 0.01 & 1.92 & 0.02 & 0.04 \\
           & $4^{+}_{1}$ & 0.24 & 1.76 & 0.01 & 0.01 & 1.96 & 0.02 & 0.01 \\
           & $3^{-}_{1}$ & 0.98 & 1.02 & 0.05 & 0.03 & 1.67 & 0.08 & 0.18 \\
           & $5^{-}_{1}$ & 1.00 & 1.00 & 0.05 & 0.04 & 1.79 & 0.07 & 0.05 \\
$^{94}$Zr: &             &      &      &      &      &      &      &      \\
           & $2^{+}_{1}$ & 0.15 & 1.85 & 0.06 & 0.01 & 3.79 & 0.06 & 0.08 \\
           & $4^{+}_{1}$ & 0.13 & 1.87 & 0.05 & 0.01 & 3.87 & 0.05 & 0.02 \\
           & $3^{-}_{1}$ & 0.99 & 1.01 & 0.08 & 0.05 & 3.52 & 0.15 & 0.20 \\
           & $5^{-}_{1}$ & 1.00 & 1.00 & 0.11 & 0.08 & 3.48 & 0.18 & 0.14 \\
$^{96}$Zr: &             &      &      &      &      &      &      &      \\
           & $2^{+}_{1}$ & 0.12 & 1.88 & 0.09 & 0.03 & 4.75 & 0.13 & 1.01 \\
           & $4^{+}_{1}$ & 0.14 & 1.86 & 0.09 & 0.04 & 4.78 & 1.02 & 0.08 \\
           & $3^{-}_{1}$ & 0.97 & 1.03 & 0.14 & 0.76 & 4.52 & 0.20 & 0.39 \\
           & $5^{-}_{1}$ & 0.98 & 1.02 & 0.16 & 0.09 & 5.21 & 0.25 & 0.29 \\
$^{98}$Zr: &             &      &      &      &      &      &      &      \\
           & $2^{+}_{1}$ & 0.10 & 1.90 & 0.12 & 0.05 & 5.76 & 1.10 & 0.97 \\
           & $4^{+}_{1}$ & 0.18 & 1.82 & 0.15 & 0.98 & 5.66 & 0.21 & 1.01 \\ 
           & $3^{-}_{1}$ & 0.99 & 1.01 & 0.15 & 0.55 & 5.55 & 1.13 & 0.63 \\
           & $5^{-}_{1}$ & 0.91 & 1.09 & 0.28 & 0.21 & 5.58 & 0.36 & 1.57 \\
\end{tabular}
\label{tab:occupation}
\end{center}
\end{table}
\begin{table}[htbp]
\begin{center}
\caption{Occupation numbers of the three lowest-lying $0^{+}$ states in 
$^{92-98}$Zr.}
\begin{tabular}{ll|ll|lllll}
   & J$^{\pi}_{i}$ & $0g_{9/2}$ & $1p_{1/2}$ & $0h_{11/2}$ & $0g_{7/2}$ &
                $1d_{5/2}$ & $1d_{3/2}$ & $2s_{1/2}$ \\
\hline
$^{92}$Zr: &             &      &      &      &      &      &      &      \\
           & $0^{+}_{1}$ & 0.33 & 1.67 & 0.06 & 0.03 & 1.81 & 0.07 & 0.03 \\
           & $0^{+}_{2}$ & 1.70 & 0.30 & 0.05 & 0.15 & 1.61 & 0.11 & 0.08 \\
           & $0^{+}_{3}$ & 1.30 & 0.70 & 0.07 & 0.14 & 0.63 & 0.06 & 1.11 \\
$^{94}$Zr: &             &      &      &      &      &      &      &      \\
           & $0^{+}_{1}$ & 0.17 & 1.83 & 0.10 & 0.04 & 3.69 & 0.11 & 0.06 \\ 
           & $0^{+}_{2}$ & 1.78 & 0.23 & 0.13 & 0.49 & 2.75 & 0.31 & 0.33 \\
           & $0^{+}_{3}$ & 0.42 & 1.58 & 0.08 & 0.06 & 2.09 & 0.12 & 1.65 \\
$^{96}$Zr: &             &      &      &      &      &      &      &      \\
           & $0^{+}_{1}$ & 0.09 & 1.91 & 0.12 & 0.03 & 5.66 & 0.12 & 0.07 \\
           & $0^{+}_{2}$ & 0.22 & 1.78 & 0.13 & 0.09 & 3.88 & 0.15 & 1.76 \\
           & $0^{+}_{3}$ & 1.76 & 0.24 & 0.19 & 0.96 & 3.86 & 0.47 & 0.52 \\
$^{98}$Zr: &             &      &      &      &      &      &      &      \\
           & $0^{+}_{1}$ & 0.09 & 1.91 & 0.15 & 0.06 & 5.76 & 0.17 & 1.87 \\
           & $0^{+}_{2}$ & 0.44 & 1.56 & 0.27 & 0.72 & 5.50 & 1.26 & 0.26 \\
           & $0^{+}_{3}$ & 1.41 & 0.59 & 0.19 & 1.46 & 4.79 & 0.98 & 0.58 \\
\end{tabular}
\label{tab:occupation0}
\end{center}
\end{table}
\begin{table}[htbp]
\begin{center}
\caption{E2 transition rates. Numbers in parentheses indicate uncertainties in 
the last digit of the quoted experimental values. The proton and neutron effective
charges are set equal to zero in column 5 and 6, respectively. All entries in 
Weisskopf units (W.u.).}
\begin{tabular}{ll|cccc}
& TRANSITION & EXP. & \multicolumn{3}{c}{CALC.} \\ \cline{4-6}
& $B(E2;J_{i}^{\pi}\longrightarrow J_{f}^{\pi})$ & & $e^{\rm eff}_{p}=1.8, e^{\rm eff}_{n}=1.5$ & $e^{\rm eff}_{p}=0.0$ & $e^{\rm eff}_{n}=0.0$ \\
\hline
$^{90}$Zr & $B(E2;2^{+}_{1}\longrightarrow 0^{+}_{1}$) & 5.37 (20)  & 1.88 & & \\
          & $B(E2;2^{+}_{1}\longrightarrow 0^{+}_{2}$) & 5.2 (10)   & 6.21 & & \\
          & $B(E2;8^{+}_{1}\longrightarrow 6^{+}_{1}$) & 2.40 (16)  & 2.57 & & \\
\hline
$^{92}$Zr & $B(E2;2^{+}_{1}\longrightarrow 0^{+}_{1}$) & 6.4 (6)    &2.97  & 2.54 & 0.02 \\
          & $B(E2;0^{+}_{2}\longrightarrow 2^{+}_{1}$) & 14.3 (5)   & 0.30 & 0.11 & 0.77 \\
          & $B(E2;4^{+}_{1}\longrightarrow 2^{+}_{1}$) & 4.04 (12)  & 0.40 & 0.30 & 0.01 \\
          & $B(E2;6^{+}_{1}\longrightarrow 4^{+}_{1}$) & $>$ 0.00098 & 0.24 & 0.002 & 0.20 \\
\hline
$^{94}$Zr & $B(E2;2^{+}_{1}\longrightarrow 0^{+}_{1}$) & 4.4 (5)    & 3.69 & 3.56 & 0.001 \\
          & $B(E2;0^{+}_{2}\longrightarrow 2^{+}_{1}$) & 9.3 (4)    & 0.01 & 0.05 & 0.11 \\
          & $B(E2;4^{+}_{1}\longrightarrow 2^{+}_{1}$) & 0.876 (23) & 1.26 & 1.22 & 0.0004 \\
\hline
$^{96}$Zr & $B(E2;2^{+}_{1}\longrightarrow 0^{+}_{1}$) & 4 (3)      & 0.05 & 0.03 & 0.002\\
          & $B(E2;2^{+}_{2}\longrightarrow 0^{+}_{1}$) & $>$ 0.020  & 0.13 & 0.10 & 0.002\\
          & $B(E2;2^{+}_{2}\longrightarrow 0^{+}_{2}$) & $>$ 2.7    & 1.17 & 1.03 & 0.005\\
\hline
$^{98}$Zr & $B(E2;2^{+}_{1}\longrightarrow 0^{+}_{1}$) & $>$ 0.24   & 0.02 & 0.01 & 0.0005 \\
          & $B(E2;2^{+}_{1}\longrightarrow 0^{+}_{2}$) & $>$ 0.04   & 0.64 & 0.48 & 0.01 \\ 
          & $B(E2;0^{+}_{3}\longrightarrow 2^{+}_{1}$) & 51 (5)     & $\approx$0.00 & 0.02 & 0.02 \\ 
\end{tabular}
\label{tab:e2}
\end{center}
\end{table}

\begin{table}[htbp]
\begin{center}
\caption{Occupation numbers in $^{97}$Zr.}
\begin{tabular}{ll|ll|lllll}
   & J$^{\pi}_{i}$ & $0g_{9/2}$ & $1p_{1/2}$ & $0h_{11/2}$ & $0g_{7/2}$ &
                $1d_{5/2}$ & $1d_{3/2}$ & $2s_{1/2}$ \\
\hline
$^{97}$Zr: & $1/2^{+}_{1}$ & 0.09 & 1.91 & 0.11 & 0.03 & 5.73 & 0.15 & 0.99 \\
           & $3/2^{+}_{1}$ & 0.10 & 1.90 & 0.11 & 0.04 & 5.72 & 1.04 & 0.08 \\
           & $5/2^{+}_{1}$ & 0.11 & 1.89 & 0.12 & 0.05 & 4.83 & 0.14 & 1.87 \\
           & $7/2^{+}_{1}$ & 0.26 & 1.74 & 0.15 & 1.00 & 5.51 & 0.17 & 0.17 \\
           & $5/2^{+}_{2}$ & 0.15 & 1.85 & 0.10 & 0.05 & 4.77 & 1.06 & 1.04 \\ 
           & $7/2^{+}_{2}$ & 0.14 & 1.86 & 0.11 & 0.05 & 4.82 & 1.04 & 0.99 \\
           & $3/2^{+}_{2}$ & 0.18 & 1.82 & 0.11 & 0.06 & 4.77 & 1.08 & 0.98 \\ 
           & $11/2^{}_{1}$ & 0.88 & 1.12 & 0.26 & 0.07 & 5.83 & 0.27 & 0.88 \\
\end{tabular}
\label{tab:occ}
\end{center}
\end{table}
\begin{table}[htbp]
\begin{center}
\caption{Theoretical and experimental energy levels in $^{91}$Zr and $^{93}$Zr. 
Energies are given in MeV.}
\begin{tabular}{cccc|cccc}
\multicolumn{4}{c}{$^{91}$Zr} & \multicolumn{4}{c}{$^{93}$Zr} \\
$J^{\pi}$ & SM & $J^{\pi}$ & EXP & $J^{\pi}$ & SM & $J^{\pi}$ & EXP \\
\hline
$5/2^{+}$  & 0.0   & $5/2^{+}$ & 0.0     & 
$5/2^{+}$  & 0.0   & $5/2^{+}$ & 0.0 \\
$1/2^{+}$  & 1.477 & $1/2^{+}$ & 1.205   & 
$3/2^{+}$  & 0.182 & $3/2^{+}$ & 0.269 \\
$5/2^{+}$  & 1.717 & $5/2^{+}$ & 1.466   & 
$9/2^{+}$  & 0.568 & $1/2^{+}$ & 0.947 \\
$7/2^{+}$  & 2.015 & $7/2^{+}$ & 1.882   & 
$1/2^{+}$  & 1.307 & $1/2^{+}$ & 1.018 \\
$9/2^{+}$  & 2.094 & $3/2^{+}$ & 2.042   & 
$7/2^{+}$  & 1.686 & $1/2^{+}$ & 1.169 \\
$3/2^{+}$  & 2.361 & $(9/2)^{+}$ & 2.131 & 
$3/2^{+}$  & 1.755 & $1/2^{+}$ & 1.222 \\
$7/2^{+}$  & 2.383 & $7/2^{+}$ & 2.201   & 
$5/2^{+}$  & 1.895 & $3/2^{+},5/2^{+}$ & 1.425 \\
$13/2^{+}$ & 2.403 & $(5/2,7/2)$ & 2.367 & 
$5/2^{+}$  & 2.114 & $(1/2^{+},3/2,5/2^{+})$ & 1.450 \\
$5/2^{+}$  & 2.405 & $3/2^{+},5/2^{+}$ & 2.535 & 
$3/2^{+}$  & 2.208 & $9/2^{+},7/2^{+}$ & 1.463 \\

$11/2^{+}$ & 2.444 & $1/2^{+}$ & 2.558 & 
$7/2^{+}$  & 2.211 & $(1/2^{+},3/2,5/2^{+})$ & 1.470 \\
\hline
$5/2^{-}$  & 1.063 & $(11/2)^{-}$ & 2.170 & 
$13/2^{-}$ & 1.417 & $(9/2^{-},11/2^{-})$ & 2.025 \\ 
$15/2^{-}$ & 1.173 & $(5/2)^{-}$ & 2.190  & 
$9/2^{-}$  & 1.445 & $(9/2^{-},11/2^{-})$ & 2.363 \\
$13/2^{-}$ & 1.423 & $(13/2)^{-}$ & 2.260 & 
$11/2^{-}$ & 1.507 & $(9/2^{-},11/2^{-})$ & 2.662 \\
$11/2^{-}$ & 1.430 & $(15/2)^{-}$ & 2.288 &
            &       &   & \\
$7/2^{-}$   & 1.520 & $(11/2)^{-}$ & 2.321 &
            &       &   & \\
\end{tabular}
\label{tab:91-93zr}
\end{center}
\end{table}
\begin{table}[htbp]
\begin{center}
\caption{Theoretical and experimental energy levels in $^{95}$Zr and $^{97}$Zr. 
Energies are given in MeV.}
\begin{tabular}{cccc|cccc}
\multicolumn{4}{c}{$^{95}$Zr} & \multicolumn{4}{c}{$^{97}$Zr} \\
$J^{\pi}$ & SM & $J^{\pi}$ & EXP &$J^{\pi}$ & SM & $J^{\pi}$ & EXP \\  
\hline
$5/2^{+}$  & 0.0   & $5/2^{+}$ & 0.0   &
$1/2^{+}$  & 0.0   &  $1/2^{+}$ & 0.0 \\ 
$1/2^{+}$  & 1.021 & $1/2^{+}$ & 0.954 &
$3/2^{+}$  & 1.062 &  $3/2^{+}$ & 1.103 \\
$3/2^{+}$  & 1.285 & $3/2^{+},5/2^{+}$ & 1.14  &  
$5/2^{+}$  & 1.168 &  $7/2^{+}$ & 1.264 \\
$7/2^{+}$  & 1.482 & $3/2^{+},5/2^{+}$ & 1.324 &
$7/2^{+}$  & 1.940 &  $(5/2^{+})$ & 1.400 \\
$5/2^{+}$  & 1.613 & $7/2^{+},9/2^{+}$ & 1.618 &
$5/2^{+}$  & 2.427 &  $(5/2^{+})$ & 1.859 \\
$9/2^{+}$  & 1.857 & $(3/2)^{+}$ & 1.618 & 
$7/2^{+}$  & 2.457 &  $(5/2^{+})$ & 1.997 \\
$3/2^{+}$  & 2.051 & $(5/2)^{+}$ & 1.722 &
$3/2^{+}$  & 2.604 &  $(3/2,5/2)$ & 2.058 \\
$7/2^{+}$  & 2.306 & $1/2^{(+)},3/2,5/2^{+}$ & 1.904 &
$9/2^{+}$  & 2.644 &  $(7/2)^{+}$ & 2.234 \\
$1/2^{+}$  & 2.414 & $1/2^{(+)},3/2,5/2^{+}$ & 1.940 &
$5/2^{+}$  & 2.759 &  $(7/2)^{+}$ & 2.508 \\
$5/2^{+}$  & 2.490 & $5/2^{(+)}$ & 1.956 &
$7/2^{+}$  & 2.786 &  $(7/2)^{+}$ & 3.161 \\
\hline
$13/2^{-}$ & 1.919 & $9/2^{-},11/2^{-}$ & 2.025 & 
$11/2^{-}$ & 2.356 & $(7/2^{-})$ & 1.807 \\
$11/2^{-}$ & 2.039 & $1/2^{-},3/2^{-}$ & 2.816 &
$9/2^{-}$  & 2.492 & $(11/2^{-})$ & 2.264 \\
\end{tabular}
\label{tab:95-97zr}
\end{center}
\end{table}
%
\begin{table}[htbp]
\begin{center}
\caption{Low-lying states in $^{90}$Zr and $^{92}$Zr. Energies are given in MeV.}
\begin{tabular}{ccc|cccc}
\multicolumn{3}{c}{$^{90}$Zr} & \multicolumn{4}{c}{$^{92}$Zr} \\
$J^{\pi}$ & $^{88}$Sr-core & EXP & $J^{\pi}$ & $^{90}$Zr-core & $^{88}$Sr-core & EXP \\
\hline
$0^{+}_{2}$ & 1.706 & 1.761 & $0^{+}_{2}$ & 2.738 & 1.693 & 1.383 \\   
$2^{+}_{1}$ & 2.003 & 2.186 & $2^{+}_{1}$ & 0.601 & 0.581 & 0.935 \\
$2^{+}_{2}$ &       & 3.309 & $2^{+}_{2}$ & 1.792 & 1.921 & 1.847 \\
            &       &       & $2^{+}_{3}$ & 3.093 & 2.167 & 2.067 \\
$3^{-}_{1}$ &       & 2.748 & $3^{-}_{1}$ & 4.018 & 1.904 & 2.340 \\ 
$4^{+}_{1}$ & 2.235 & 3.077 & $4^{+}_{1}$ & 0.850 & 0.823 & 1.496 \\
$4^{-}_{1}$ & 1.693 & 2.739 & $4^{+}_{2}$ & 2.677 & 2.379 & 2.398 \\ 
$5^{-}_{1}$ &       & 2.319 & $5^{-}_{1}$ & 4.261 & 1.316 & 2.486 \\ 
$6^{+}_{1}$ & 2.317 & 3.448 & $6^{+}_{1}$ & 3.206 & 2.596 &       \\
\end{tabular}
\label{tab:90-92zr}
\end{center}
\end{table}
\begin{table}[htbp]
\begin{center}
\caption{Low-lying states in $^{94}$Zr and $^{96}$Zr. Energies are 
given in MeV.}
\begin{tabular}{cccc|cccc}
\multicolumn{4}{c}{$^{94}$Zr} & \multicolumn{4}{c}{$^{96}$Zr} \\
$J^{\pi}$ & $^{90}$Zr-core & $^{88}$Sr-core & EXP & $J^{\pi}$ & $^{90}$Zr-core &$^{88}$Sr-core & EXP \\
\hline
$0^{+}_{2}$ & 2.254 & 2.213 & 1.300 & $0^{+}_{2}$ & 1.602 & 2.228 & 1.582 \\
$2^{+}_{1}$ & 0.557 & 0.520 & 0.919 & $2^{+}_{1}$ & 1.097 & 1.426 & 1.751 \\  
$2^{+}_{2}$ & 1.490 & 1.764 & 1.671 & $2^{+}_{2}$ & 2.105 & 2.661 & 2.226 \\
$2^{+}_{3}$ & 1.930 & 2.228 & 2.151 & $2^{+}_{3}$ & 2.211 & 2.768 & 2.669 \\
$3^{-}_{1}$ & 3.923 & 2.223 & 2.058 & $3^{-}_{1}$ & 3.833 & 3.732 & 1.897 \\
$4^{+}_{1}$ & 0.889 & 0.817 & 1.470 & $4^{+}_{1}$ & 2.054 & 2.528 & 2.750 \\
$4^{+}_{2}$ & 1.898 & 2.162 & 2.330 & $4^{+}_{2}$ & 2.230 & 2.869 & 2.857 \\
$5^{-}_{1}$ & 4.200 & 2.673 & 2.945 & $5^{-}_{1}$ & 4.141 & 2.306 & 3.120 \\
\end{tabular}
\label{tab:94-96zr}
\end{center}
\end{table}
\begin{table}[htbp]
\begin{center}
\caption{Low-lying states in $^{98}$Zr. Energies are given in MeV.}
\begin{tabular}{ccccc}
\multicolumn{5}{c}{$^{98}$Zr} \\
$J^{\pi}$ & $^{90}$Zr-core & $^{88}$Sr-core & $^{94}$Sr-core & EXP \\
\hline
$0^{+}_{2}$ & 1.641 & 1.904 & 0.529 & 0.854 \\ 
$0^{+}_{3}$ & 2.773 & 2.471 & 1.969 & 1.859 \\
$2^{+}_{1}$ & 1.300 & 1.463 & 1.152 & 1.223 \\ 
$2^{+}_{2}$ & 2.207 & 2.619 & 1.773 & 1.591 \\ 
$2^{+}_{3}$ & 2.442 &       & 2.149 & 1.744 \\ 
$3^{-}_{1}$ & 3.983 & 3.579 & 2.597 & 1.806 \\ 
$4^{+}_{1}$ & 2.147 & 2.449 & 1.574 & 1.843 \\ 
$4^{+}_{2}$ & 2.048 & 2.650 & 1.918 & 2.330 \\ 
$5^{-}_{1}$ & 3.576 & 3.478 & 1.318 & 2.800 \\ 
\end{tabular}
\label{tab:98zr}
\end{center}
\end{table}

\begin{thebibliography}{200}
\bibitem{schu80} F.\ Schussler, J.A.\ Pinston, E.\ Monnand, A.\
                 Moussa, G.\ Jung, E.\ Koglin, B.\ Pfeiffer,
                 R.V.F. Janssens and J.\ van Klinken, Nucl.\ Phys.\ 
                 A {\bf 339}, 415 (1980).
\bibitem{beck84} K.\ Becker, G.\ Jung, K.-H.\ Kobras, H.\ Wollnik and B.\
                  Pfeiffer, Z.\ Phys.\ A {\bf 319}, 193 (1984).
\bibitem{mach89} H.\ Mach, M.\ Moszy\`{n}ski, R.L.\ Gill, J.A.\ Winger,
                 John C.\ Hill, G.\ Mol\'{n}ar and K.\ Sistemich, Phys.\ 
                 Lett.\ B {\bf 230}, 21 (1989).
\bibitem{lher97} G.\ Lhersonneau {\em et al.}, Phys.\ Rev.\ C {\bf 56},1445 (1997).
\bibitem{lher94} G.\ Lhersonneau {\em et al.}, Phys.\ Rev.\ C{\bf 49}, 1379 (1994);
                  G.\ Lhersonneau {\em et al.}, Phys.\ Rev.\ C {\bf 54},1117 (1996).
\bibitem{gloe75} D.H.\ Gloeckner, Phys.\ Lett.\ B {\bf 42}, 381
                 (1972); D.H.\ Gloeckner, Nucl.\ Phys.\ A {\bf 253}, 301 (1975). 
\bibitem{hals93} P.\ Halse, J.\ Phys.\ G {\bf 19},1859 (1993).
\bibitem{ips75} S.S.\ Ipson, K.C.\ McLean, W.\ Booth, J.G.B.\ Haigh and
                R.N.\ Glover, Nucl.\ Phys.\ A {\bf 253}, 189 (1975).
\bibitem{auer65} N.\ Auerbach and I.\ Talmi, Nucl.\ Phys.\ {\bf 64}, 458 (1965).
\bibitem{eng93} T.\ Engeland, M.\ Hjorth-Jensen, A.\ Holt and E.\
                Osnes, Phys.\ Rev.\ C {\bf 48}, 535 (1993).
\bibitem{holt97} A.\ Holt, T.\ Engeland, M.\ Hjorth-Jensen, E.\ Osnes and
                 J.\ Suhonen, Nucl.\ Phys.\ A {\bf 618}, 107 (1997).
\bibitem{suh98} J.\ Suhonen, J.\ Toivanen, A.\ Holt, T.\ Engeland, 
                M.\ Hjorth-Jensen and E.\ Osnes, Nucl.\ Phys.\ A {\bf
                628}, 41 (1998).
\bibitem{holt98} A.\ Holt, T.\ Engeland, M.\ Hjorth-Jensen and E.\ Osnes, 
                 Nucl.\ Phys.\ A {\bf 634}, 41 (1998).
\bibitem{hko95} M.\ Hjorth-Jensen, T.T.S.\ Kuo and E.\ Osnes, Phys.\
                Reports {\bf 261}, 125 (1995).
\bibitem{cdbonn} R.\ Machleidt, F.\ Sammarruca and Y.\ Song,
                 Phys.\ Rev.\  C {\bf 53}, 1483 (1996)
\bibitem{nim} V.G.J.\ Stoks, R.A.M.\ Klomp, C.P.F. Terheggen and J.J.\
              de Swart, Phys.\ Rev.\ C {\bf 48}, 792 (1993).
\bibitem{v18} R.B.\ Wiringa, V.G.J.\ Stoks and R.\ Schiavilla, Phys.\ Rev.\
              C {\bf 51}, 38 (1995).
\bibitem{mac89}  R.\ Machleidt, Adv.\ Nucl.\ Phys.\ {\bf 19}, 189 (1989). 
\bibitem{ko90}  T.T.S.\ Kuo and E.\ Osnes, Folded-Diagram Theory of the
                Effective Interaction in Atomic Nuclei, Springer
                Lecture Notes in Physics, (Springer, Berlin, 1990) Vol.\ 364.
\bibitem{whi77} R.R.\ Whitehead, A.\ Watt, B.J.\ Cole and I.\ Morrison, Adv.\
                Nucl.\ Phys.\ {\bf 9}, 123 (1977). 
\bibitem{eng95} T.\ Engeland, M.\ Hjorth-Jensen, A.\ Holt and E.\ Osnes, 
                Phys.\ Scr.\ T {\bf 56}, 58 (1995).
\bibitem{saga84} A.\ Saganek, V.\ Meyer, S.\ Mirowski, M.\ Oteski, 
                 M.\ Sieme\'{a}ski, E.\ Wesolowski and Z.\ Wilhelmi,
                 J.\ Phys.\ G {\bf 10}, 549 (1984).
\bibitem{clea78} T.P.\ Cleary, Nucl.\ Phys.\ A {\bf 301}, 317 (1978).
\bibitem{rama73} S.\ Ramavataram, B.\ Goulard and J.\ Bergeron, Nucl.\ Phys.\
                A {\bf 207}, 140 (1973).
\bibitem{hoff74} P.\ Hoffmann-Pinther and J.L.\ Adams, Nucl.\ Phys.\
                 A {\bf 229}, 365 (1974).
\bibitem{chuu79} D.S.\ Chuu, M.M.\ King Yen, Y.\ Shan and T.\ Hsien, Nucl.\
                 Phys.\ A {\bf 321}, 415 (1979).
\bibitem{saha79} A.\ Saha, G.D.\ Jones, L.W.\ Put and R.H.\ Siemssen, 
                 Phys.\ Lett.\ B {\bf 82}, 208 (1979). 
\bibitem{brow76} B.A.\ Brown, P.M.S.\ Lesser and D.B.\ Fossan, Phys.\ Rev.\ 
                 C {\bf 13} (1976) 1900.
\bibitem{hjor99} M.\ Hjorth-Jensen, unpublished.
\bibitem{sche80} L.\ Schellenberg, B.\ Robert-Tissot, K.\ Kaeser, L.A.\ 
                 Schaller, H.\ Schneuwly, G.\ Fricke, S.\ Gluckert, G.\ Mallot 
                 and E.B.\ Shere, Nucl.\ Phys.\ A {\bf 333} (1980) 333.
\bibitem{zuk94}  A.P.\ Zuker, Nucl.\ Phys.\ A {\bf 576}, 65 (1994); G.\ 
                 Martinez-Pinedo, A.P.\ Zuker, A.\ Poves and E.\
                 Caurier, Phys.\ Rev.\ C {\bf 55}, 187 (1997).
\end{thebibliography}
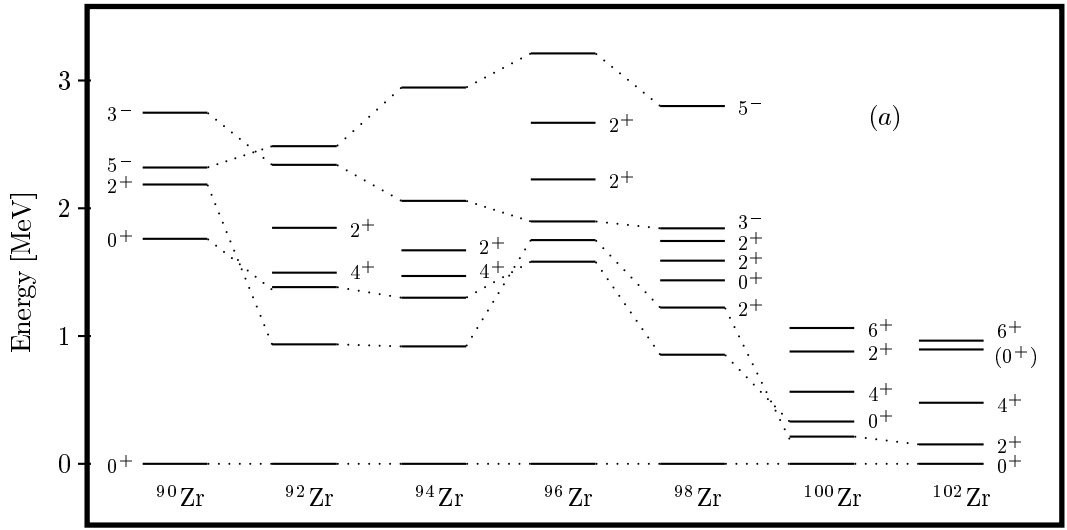
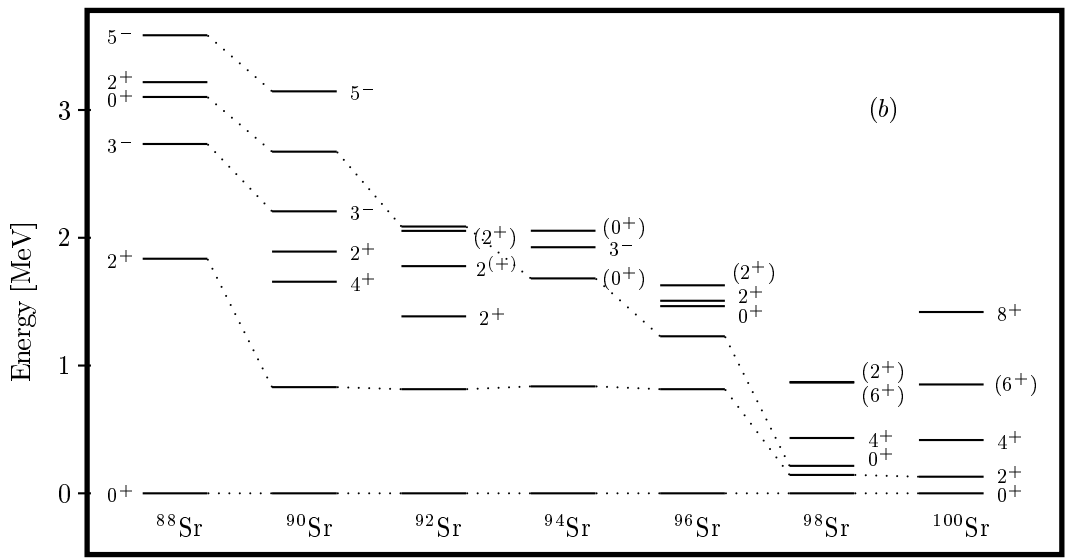
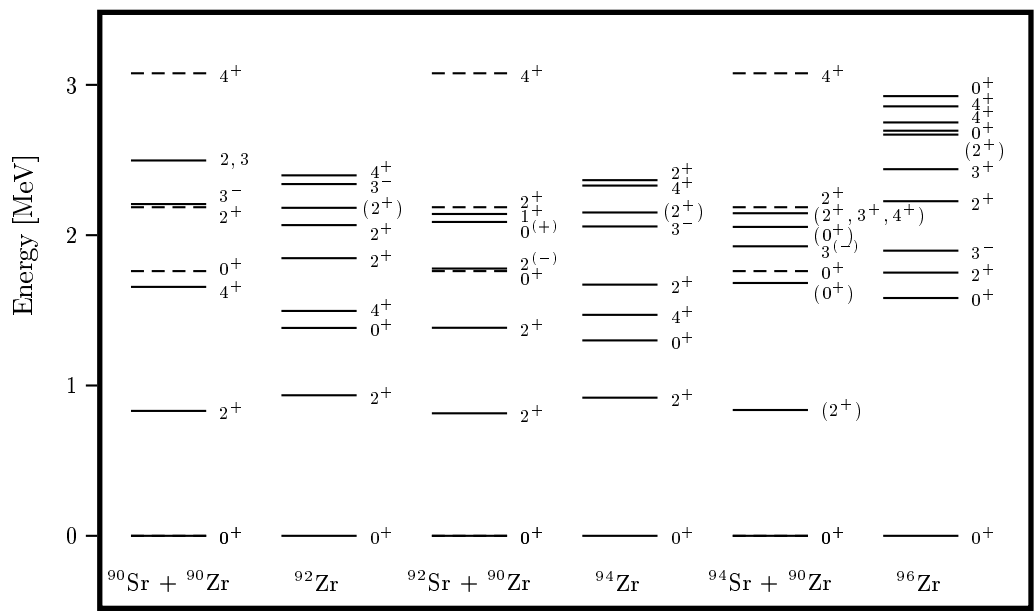
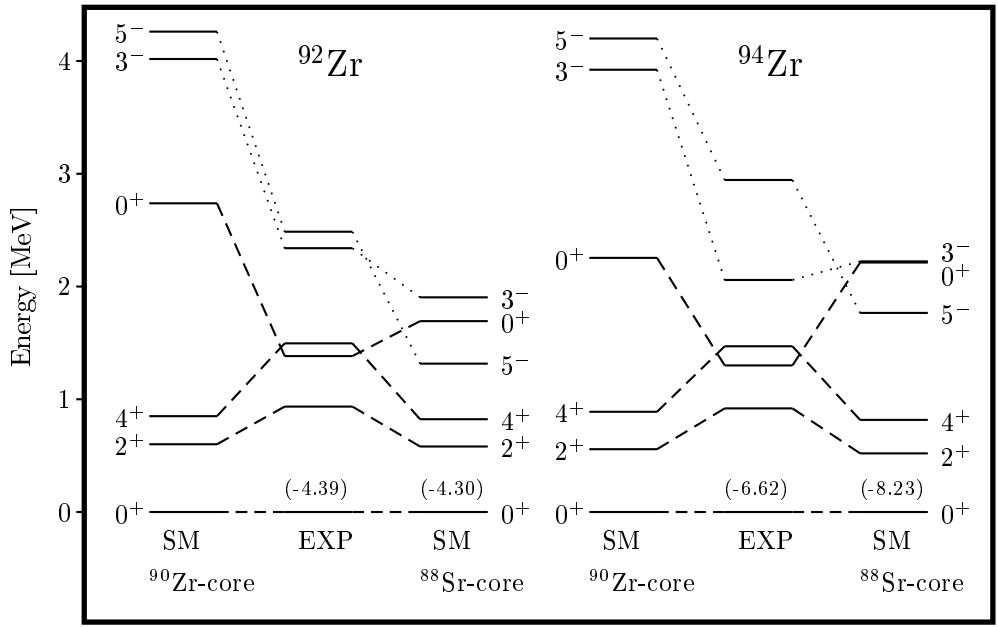
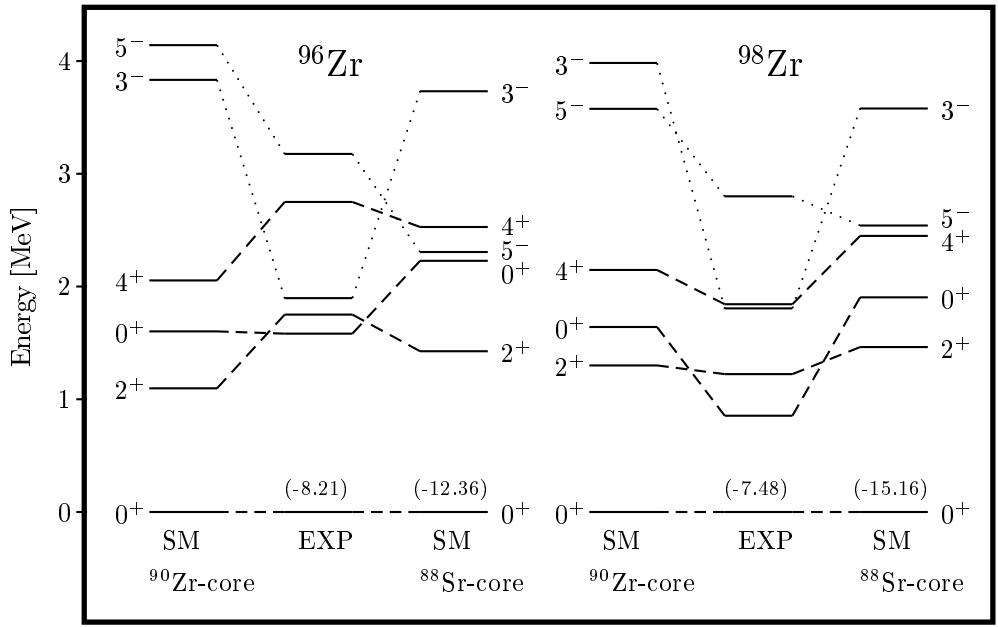
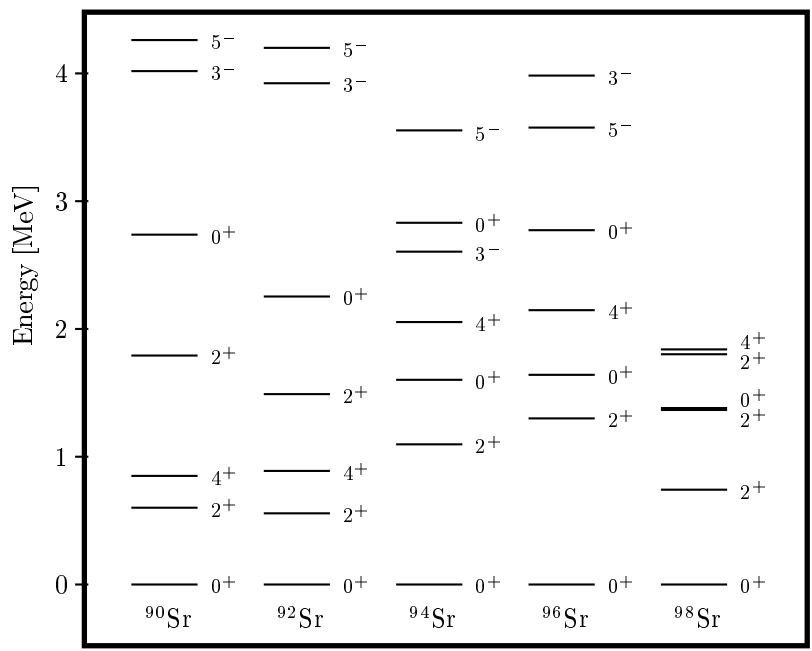
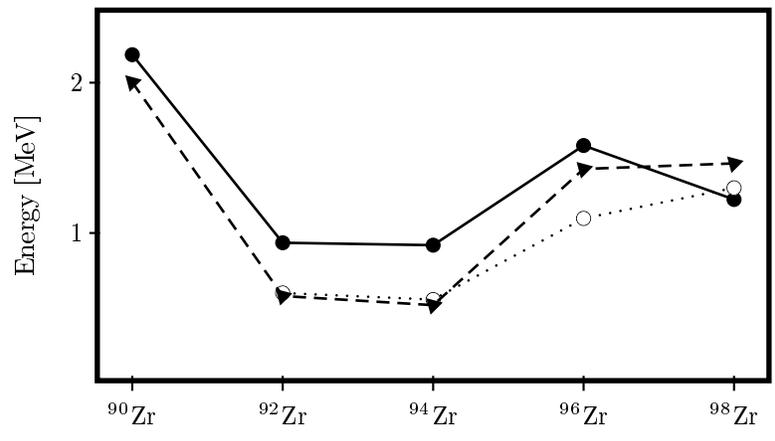
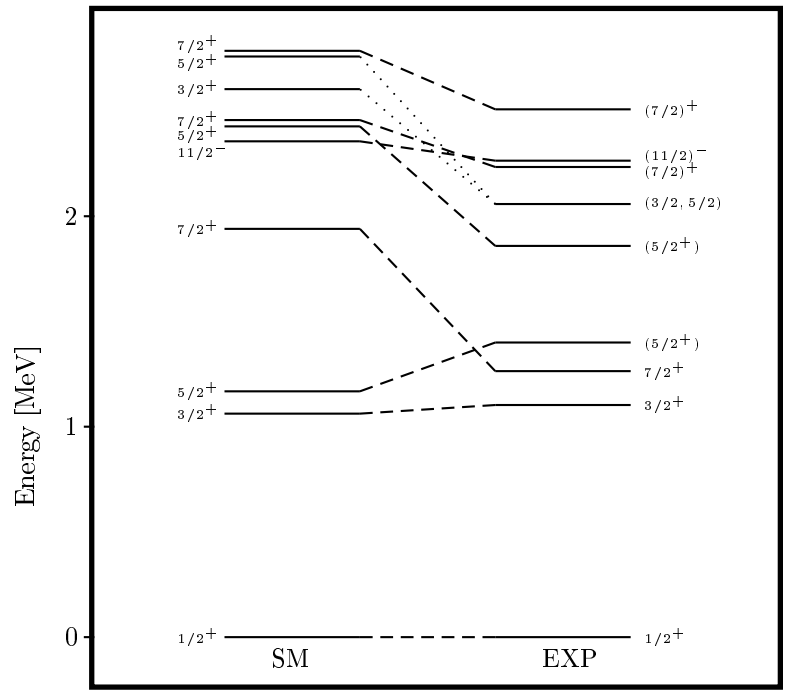
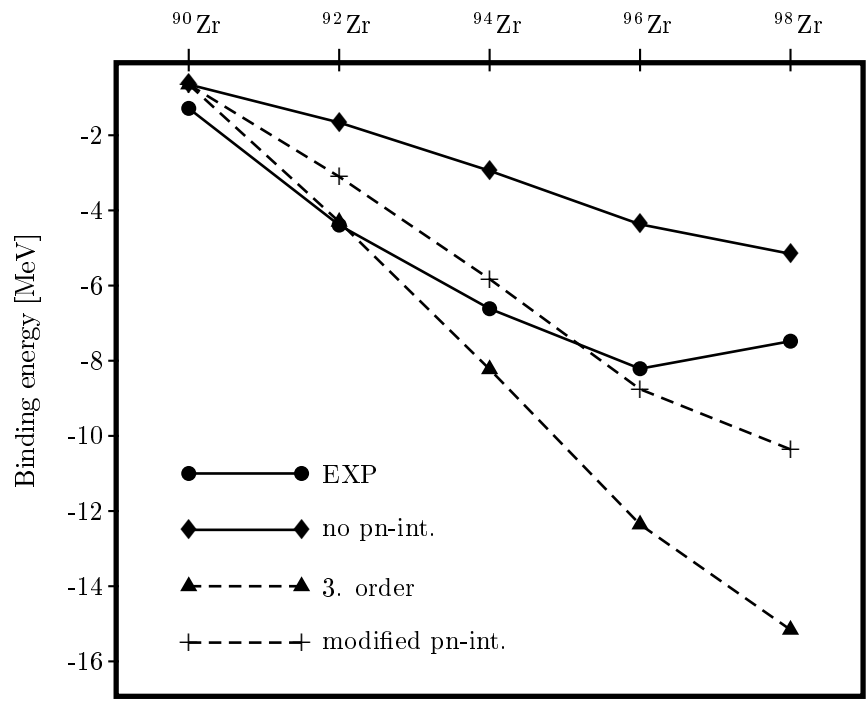
\end{document}